\documentclass[11pt,twoside]{article}

\usepackage{asp2006}
\usepackage{epsf}
\usepackage{psfig}
\usepackage{lscape}
\usepackage{graphicx}

\begin{document}
\title{The Radio Sky on Short Timescales with LOFAR: \\ Pulsars and Fast Transients} 
\author{J.W.T. Hessels$^{1,2}$, B.W. Stappers$^3$, \& J. van Leeuwen$^{1,2}$ \\ 
on behalf of the LOFAR Transients Key Science Project} 

\affil{$^1$ASTRON, The Netherlands\\  
$^2$University of Amsterdam, The Netherlands \\
$^3$University of Manchester, United Kingdom}

\begin{abstract} 
LOFAR, the ``low-frequency array'', will be one of the first in a new
generation of radio telescopes and Square Kilometer Array (SKA)
pathfinders that are highly flexible in capability because they are
largely software driven.  LOFAR will not only open up a mostly
unexplored spectral window, the lowest frequency radio light
observable from the Earth's surface, but it will also be an
unprecented tool with which to monitor the transient radio sky over a
large field of view and down to timescales of milliseconds or less.
Here we discuss LOFAR's current and upcoming capabilities for
observing fast transients and pulsars, and briefly present recent
commissioning observations of known pulsars.
\end{abstract}


\section{Monitoring the (Low-Frequency) Radio Sky at High Time Resolution}   

A key factor in successfully studying the dynamic and explosive events
associated with compact objects is the ability to monitor a large area
of the sky continuously for transient sources.  This has been done for
many years now, and with great success in, e.g., X-ray astronomy,
where ``all-sky'' monitors onboard a number of space-based telescopes
can alert the community to exotic transient events and trigger
directed, multi-wavelength observations.  The ability to do similar,
sensitive all-sky monitoring in the radio regime would open up a
largely unexplored domain: ``the transient radio sky'' \citep[see,
e.g.,][]{clm04}.  Despite the strong scientific motivation to
characterize radio transients on a wide-range of timescales, this
exploration is hampered by technical challenges intrinsic to radio
astronomical observations.  The observational requirements are well
summarized by the following simple figure of merit (FoM), which
applies in general to any transient survey\footnote{Here we choose to
give an extra weighting to $A_{eff}$.  One may also consider a FoM that
scales linearly with $A_{eff}$.} \citep[see also][and references therein
for a deeper discussion of survey metrics]{clm04,cor08}:

\begin{equation}
FoM \propto A^2_{eff} \frac{\Omega}{\Delta \Omega} \frac{T}{\Delta T}
\end{equation}

To effectively probe transients over a wide range of source parameter
space, including faint and rare events, one must maximize this FoM and
hence also maximize: $A_{eff}$, the effective collecting area,
i.e. raw sensitivity, of the detector; $\Omega$, the instantaneous
field of view (FoV); and $T$, the total time spent observing the sky.
At the same time, one should also maintain adequate spatial ($\Delta
\Omega$) and time ($\Delta T$) resolution to provide reasonable source
localization (crucial for multi-wavelength follow-up and
identification) and to resolve short timescale phenomena.

Traditional ``single pixel'' large single dish radio telescopes and
standard radio interferometers fall short of providing a good FoM
because they cannot simultaneously provide high sensitivity and large
FoV.  One way to achieve both of these requirements is to combine the
signals of many small elements, each with close to ``all-sky'' FoV, to
produce multiple, sensitive beams on the sky.  Such a telescope makes
use of large computing resources to do more with the signals received
by each element, and could appropriately be dubbed a ``software
telescope''.  Such instruments are important pathfinders to the Square
Kilometer Array (SKA), which will ultimately provide a monumental leap
in our ability to monitor the radio sky in real time
\citep[e.g.][]{cor08}.  Here we discuss specifically the low-frequency
array (LOFAR), which incorporates many innovative techniques and will
itself revolutionize our ability to observe the transient radio sky.

Perhaps the regime in which the transient radio sky has been least
well characterized is on timescales of seconds and shorter.  These
timescales are too short to be probed satisfactorily by standard
imaging techniques, but can be well studied with beam-formed
timeseries, i.e. ``pulsar-like'', data.  LOFAR will provide both
pulsar-like and imaging data, to a certain degree simultaneously.  In
addition to opening a new window on the transient radio sky, LOFAR
will probe the relatively unexplored spectral window of $\sim 30 -
240$\,MHz, the lowest radio frequencies observable from the Earth's
surface.

This low observing frequency presents its own opportunities, but also
difficult challenges for observing ``fast'' transients (timescale $<
1$\,s) and pulsars.  Ultimately, the effective time resolution of
observations is limited by propagation effects in the interstellar
medium such as scattering and dispersion, which become very strong at
low frequency.  Impulsive radio frequency interference (RFI) is also a
major concern for time-domain based searches, although it can be
mitigated by a number of signal processing techniques and the use of a
many-element interferometer.

Despite these challenges, the potential scientific rewards of probing
new parameter space(s) are enticing.  In addition to targetted
observations of known pulsars to study, for example, their single
pulse properties \citep{slk+07}, an all-sky survey for pulsars and
fast transients will be done \citep[see][for a survey
simulation]{ls08}.  This survey will discover potentially hundreds of
new pulsars, and will likely detect the majority of the nearby ($d <
2$\,kpc) radio emitting neutron stars, including those that show
transient behaviour \citep[e.g. the ``rotating radio
transients''][]{mll+06}.  At the same time, this survey is also sensitive
to other types of sources showing (dispersed) sub-second bursts, for
instance nearby exoplanets and flare stars, but potentially new source
classes as well.  A key advantage of such a LOFAR survey is that its
large FoV allows for potentially long dwell times when doing an all
sky survey, which is particularly important for identifying
transients, especially rare events.  Coupled with an imaging ``radio sky
monitor'' \citep{fws+08}, LOFAR will repeatedly cover the sky to detect
even extremely rare transient events with timescales of nano-seconds
to years.

\section{LOFAR Observing Modes for Pulsars and Fast Transients}   

The LOFAR telescope\footnote{For a broader description of the LOFAR
system and a status update, see also de Bruyn et al. 2009 in these
proceedings and references therein.} is actually two separate arrays,
operating respectively from $30-80$\,MHz (the Low-Band Antennae, or
LBAs, which consist of single dipole antennae) and $100-240$\,MHz (the
High-Band Antennae, or HBAs, which consist of integrated tiles of 16
phased dipoles each).  These antennae are grouped into several dozen
stations of $24-96$ elements each, and will soon be spread over the
Netherlands and neighbouring countries.  The core of LOFAR has the
highest filling factor (i.e. antennae per unit land area) in the array
and provides the best compromise between raw sensitivity and FoV for
most transient and pulsar searches.  The first incarnation of LOFAR
will have 36 core stations of 24 HBA tiles each, roughly 50\% of the
total collecting area of the array.  Crucial to LOFAR's operation is
the Blue Gene P super-computer (hereafter BG/P) which combines the
signals from the stations to produce either visibilities for
interferometric imaging or (coherent) beams from combined elements.
The fact that so much of LOFAR's low-level signal processing is
handled real time in software makes it a highly flexible system.  In
many ways, the system is more limited by the available processing
power and the amount of data that can be transported back from the
stations for processing than it is by the antennae themselves.  Here
we summarize the LOFAR observational modes, also largely applicable to
other similar telescopes, most relevant to observing transients on
short timescales.

{\bf (Multiple) Station Beams}: Independent of BG/P, each LOFAR
station is capable of forming a ``station beam'', which is a coherent
addition of all the elements within a station.  Up to 8 independently
pointed station beams can be formed simultaneously within the element
primary beam pattern, each having a maximum total bandwidth of
$32/n_{beams}$\,MHz.  An advantage is that the observing bandwidth
need not be contiguous, but rather can be spread over the entire
available receiver band, avoiding areas of the spectrum that are
consistently contaminated by strong RFI.  

This simultaneous coverage over a fractional bandwidth of roughly 1/2
also provides the exciting possibility of catching fast transients in
the act: identifying these in real time, first at the top of the band,
and then adjusting the system to follow the source as it appears later
at lower frequency.  For instance, this could happen because of
interstellar dispersion, which can easily cause a delay of several
minutes to an hour at these low observing frequencies, compared with
higher observing frequency.

One can imagine using the LOFAR stations separately, either single
stations for projects requiring less collecting area but more total
observing time or as a collective, each station covering up to 8
independent regions of the sky in a mode similar to the ``Fly's Eye''
experiment running on the Allen Telescope Array \citep{sie08}.  Since
each station still has a reasonable collecting area, comparable to
that of a $\sim 35$-m single dish telescope, this mode is interesting
for achieving the maximum possible sky coverage for relatively bright,
but rare, events (Table~\ref{tab1}).  Note that at $160$\,MHz the
largest possible FoV in this mode is still only about 9\% of the
entire sky at any given time (nonetheless a major leap forward
compared with previous radio telescopes).  This increases to roughly
full hemispheral coverage at 40\,MHz but, at this very low frequency,
propagation effects like scattering are a very severe limitation to
observing fast transients.

{\bf Incoherent Station Summation}: To increase sensitivity, the
signals from individual stations must be combined.  An incoherent
station summation corrects for the rough time, but not the phase
delays between stations when observing in a particular direction
(i.e. the signals from individual stations are first converted to
powers and then summed).  The net gain in sensitivity over that of a
single station is in theory $\sqrt{n_{stations}}$, but can be better
than this in practice because of better robustness to RFI.  This
summation is computationally inexpensive and has a comparatively low
data rate (Table~\ref{tab1}).  It should furthermore be possible to
run this mode piggy-backing on other (imaging) observations.  This
provides for a potentially enormouse total time $T$ spent on the sky.
The drawback is that the instantaneous sky coverage is small compared
with the Fly's Eye mode mentioned above, and the size of the beams
gives relatively poor positional constraints compared with a coherent
addition of stations.

{\bf Coherent Station Summation}: The maximum raw sensitivity comes
from a {\it coherent} addition of the stations, in which appropriate
phase corrections are applied before summing individual station beams.
Since the beam size is now set by the longest baseline between the
stations, this results in a comparatively narrow field of view.  This
is highly desirable for constraining the position of a new source, but
precludes monitoring large portions of sky simultaneously.  To
increase FoV, one may synthesize many, potentially hundreds, of these
``pencil beams'' in order to cover as much of the station primary beam
as possible.  This mode provides both the highest achievable raw
sensitivity and excellent positional information (Table~\ref{tab1}).
The primary drawback is that the data rate is {\it extremely} high and
ultimately limits how often one can observe in this mode without first
reducing the data to a much smaller set of analysis products.  An
additional possibility is to combine only the 12 stations contained
within the 300-m LOFAR inner core, referred to collectively as the
``Superstation''.  This mode provides a nice compromise between raw
sensitivity, FoV, and spatial resolution.

\begin{table}
\begin{center}
\caption{Comparing LOFAR Beam-Formed Observing Modes}
\label{tab1}
\begin{footnotesize}
\begin{tabular}{l c c c c c}
\hline \hline
Mode               & Sensitivity & FoV        & Resolution & Data Rate & FoM     \\
                   & (Norm.)     & (sq. deg.) & (deg)      & (TB/hr)   & (Norm.) \\
\hline
Single Station     & 1.0 / 0.4   & 12.5 / 100   & 2    & 0.23 & 1.0 / 1.3 \\
Fly's Eye          & 1.0 / 0.4   & 450  / 3600  & 2    & 8.3  & 36 / 46 \\
Incoherent Sum     & 6.0 / 2.1   & 12.5 / 100   & 2    & 0.23 & 36 / 35 \\
Superstation       & 12  / 4.2   & 9.0  / 72    & 0.2  & 23   & 1040 / 1020 \\
Coherent Sum       & 36  / 13    & 0.2  / 1.6   & 0.03 & 23   & 1380 / 1440 \\
\hline \hline
\end{tabular}
\end{footnotesize}
\end{center}

{\footnotesize Sensitivities and FoMs have been normalized to that of
a single 24-tile HBA station using a single beam of bandwidth 32\,MHz.
FoV and resolution are at 160\,MHz. Quantities are quoted assuming one
beam per station (32\,MHz bandwidth) and 8 beams per station (4\,MHz
bandwidth per beam) respectively.  All modes with the exception of
``Single Station'' assume 36 core stations of 24 HBA tiles are being
included and can be recorded separately if desired.  For the
``Superstation'' and ``Coherent Sum'' modes, we assume that 100 pencil
beams can be synthesized, and that the maximum baseline between
stations is 300\,m and 2000\,m respectively.  The dwell time used in
each mode is assumed to be the same, though this would likely differ
in practice.  The data rates assume 16-bit samples, summed in
polarization, and at the maximum possible spectral/time resolution,
which for certain applications can be downgraded by a factor of a few
in order to save on disk space and processing load.}

\end{table}

Table~\ref{tab1} summarizes and quantitatively compares these possible
modes.  There is no single mode which is best for {\it all}
conceivable types of transients - for instance a mode with large
instantaneous sky coverage but comparatively low raw sensitivity will
be good for rare, bright transients but not for weaker sources.  The
use of multiple modes, covering complementary portions of transient
parameter space, is thus advisable.  Furthermore, one may also prefer
to consider the FoM per TB of recorded data, especially if one is more
limited by available processing power for offline analysis than by
observing time.  In such a case, the modes with very high data rates
become less desirable, despite their much larger nominal FoM.

\section{Commissioning Observations, Current Status, \& Prospects for the Near Future}   

\begin{figure}
\begin{center}
\includegraphics[angle=180,width=5in]{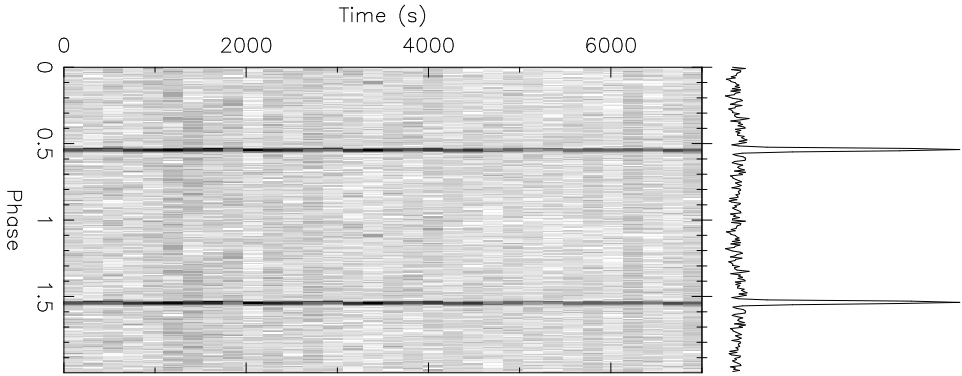}
\caption{A roughly 2-hr LOFAR observation of PSR B1919+21 ($P_{spin} =
  1.34$\,s, $DM = 12$\,pc cm$^{-3}$) from 2008 Nov 19.  The signals
  from the 4 existing HBA tiles were combined at the station itself
  into a station beam and 46 subbands of 195.3125\,kHz each were sent
  in real time via fibre back to the BG/P in Groningen.  These subbands
  formed a contiguous bandwidth of 9\,MHz, centred at 155\,MHz.
  Folding and dedispersion (optionally coherent dedispersion) was
  performed offline on the recorded data.}
\label{lofar.fig}
\end{center}
\end{figure}

The LOFAR Pulsar Working Group is part of the larger, and more
all encompassing, LOFAR Transients Key Science Project \citep[see][for
a list of collaborators and basic science case]{fws+08}, and is
actively commissioning the telescope for pulsar and fast transient
observations (with of course the crucial help of several developers
and support scientists).  A variety of observational considerations,
including dispersion, scattering, spectral turn over, and sky
background, make LOFAR's high band ($\sim 100-240$\,MHz) preferable
for most of the observations we are planning (note however that some
transient sources, like planets, may only be observable in the low
band).  Our current test bed for high-band observations consists
primarily of 4 HBA tiles, located at the LOFAR core site between Exloo
and Buinen in the Dutch province of Drenthe.  A number of
observational milestones have already been achieved, including:

\begin{itemize}

\item Detection of integrated emission from several bright known
pulsars, e.g. PSRs B0329+54, B1133+16, B1919+21, and B0809+74.  These
are being used as test sources to commission LOFAR beam-formed modes.

\item ``Blind'' detection of dispersed single pulses from PSR
B0329+54.  This same technique is sensitive to general, short
timescale, dispersed bursts.

\item Multi-hour tracking observations far from the zenith, using the
beam formed at station level (Figure~\ref{lofar.fig}).

\item Coherent addition of multiple ``stations'' - in this case the 4
HBA tiles recorded separately - into a tied-array beam.  Unlike the
station beams themselves, this beam is formed on the BG/P.

\end{itemize}

We note that 4 HBA tiles constitute only 1/6 of a core station, and
that the LOFAR core will, within the next two years, contain roughly
200 times as much collecting area as used in the current pulsar
commissioning observations.  For instance, we expect that the LOFAR
core will have enough sensitivity to detect high S/N single pulses
from roughly half of the known pulsars in the northern hemisphere.
The functionality that is currently being implemented can be scaled
relatively easily to meet this increase in collecting area.  We look
forward to making the first full-station observations in 2009.

\acknowledgements 

The LOFAR telescope is being made possible by the hard work of dozens
of engineers, technicians, developers, observers, and support
scientists.  In connection with this work, we would like to thank a
few in particular: Joe Masters, Ramesh Karuppusamy, Jan David Mol,
Michiel Brentjens, Jurjen Sluman, Geert Kuper, and Yuan Tang.


\end{document}